\documentclass[12pt,preprint]{aastex}

\begin{document}

\def\lya{\ifmmode {{\rm Ly}\alpha}\else
        Lyman-$\alpha$\fi}
\newcommand{\eqw}{\hbox{EW}}
\def\erg{\hbox{erg}}
\def\cm{\hbox{cm}}
\def\sec{\hbox{s}}
\def\f17{f_{17}}
\def\kpc{\hbox{kpc}}
\def\Mpc{\hbox{Mpc}}
\def\pMpc{\hbox{pMpc}}
\def\cMpc{\hbox{cMpc}}
\def\nm{\hbox{nm}}
\def\km{\hbox{km}}
\def\kms{\hbox{km s$^{-1}$}}
\def\year{\hbox{yr}}
\def\deg{\hbox{deg}}
\def\arcsec{\hbox{arcsec}}
\def\microJy{\mu\hbox{Jy}}
\def\zre{z_r}
\def\fesc{f_{\rm esc}}

\def\ergcm2s{\ifmmode {\rm\,erg\,cm^{-2}\,s^{-1}}\else
                ${\rm\,ergs\,cm^{-2}\,s^{-1}}$\fi}
\def\ergsec{\ifmmode {\rm\,erg\,s^{-1}}\else
                ${\rm\,ergs\,s^{-1}}$\fi}
\def\kmsMpc{\ifmmode {\rm\,km\,s^{-1}\,Mpc^{-1}}\else
                ${\rm\,km\,s^{-1}\,Mpc^{-1}}$\fi}
\def\nv{\ion{N}{5} $\lambda$1240}
\def\civ{\ion{C}{4} $\lambda$1549}
\def\oii{[\ion{O}{2}] $\lambda$3727}
\def\oiipair{[\ion{O}{2}] $\lambda \lambda$3726,3729}
\def\oiii{[\ion{O}{3}] $\lambda$5007}
\def\oiiipair{[\ion{O}{3}] $\lambda \lambda$4959,5007}
\def\tlya{\tau_\lya}
\def\llya{\ifmmode{L_\lya}\else
  $L_\lya$ \fi}
\def\sqdeg{\hbox{sq.deg}}
\def\ss{ionized bubble}

\title{Testing the Topology of Reionization}

\author{James E. Rhoads \altaffilmark{1}}

\begin{abstract}
The central overlap phase of cosmological hydrogen reionization is
fundamentally a change in the topology of ionized regions.  Before
overlap, ionized bubbles grew in isolation.  During overlap, they
merge into a percolating ionized medium, which fills an
ever-increasing volume and eventually replaces neutral gas throughout
the intergalactic medium.  Overlap can therefore be well studied using
topological statistics, and in particular the genus number of the
neutral-ionized interface. 
The most promising observational tools for applying such
tests are (a) \lya\ galaxies, and (b) 21 cm tomography.
\lya\ galaxies will be detected whenever they inhabit bubbles with
sizes $\ga 1 \pMpc$, and their presence can therefore be used to map
such bubbles.  Such large bubbles are expected during the overlap
phase, and moreover, each one should contain a few detectably bright
\lya\ galaxies.  The 21cm line in principle affords better spatial
resolution, but the required sensitivity and foreground subtraction
may be an issue.  Upcoming \lya\ surveys in the near-infrared could
thus provide our first look at the topology of reionization.
\end{abstract}

\altaffiltext{1}{School of Earth and Space Exploration,
Arizona State University, Tempe, AZ}


\section{Introduction}
The hydrogen that fills the universe became neutral at
the epoch of recombination ($z\sim 1100$).  It remained so until the
first luminous objects turned on, at perhaps $z \sim 20$.
Then, ultraviolet light from the first stars, galaxies, and
quasars began to carve out ionized bubbles in the neutral gas.  At
first, these bubbles grew in isolation.  The central period of
the reionization process came later, when the bubbles began to
overlap.  

The overlap phase is fundamentally a change in the topology of the
ionized regions.  Thus, topological tests may provide the best
diagnotics of when and how the overlap phase of reionization actually
occurred (Rhoads 2003).
Indeed,  {\it defining} overlap as the moment when 
the ionized regions of the intergalactic medium become
multiply connected provides a robust criterion for assessing the
progress of reionization.  Clustering of galaxies can result in bubbles
containing multiple ionizing sources even at very early times,
and recent work has highlighted the importance of such effects
on ionized bubble sizes (Wyithe \& Loeb 2005; Furlanetto, Zaldarriaga,
\& Hernquist 2004; Furlanetto \& Oh 2005).
Still, many bubbles must form independently in
the early phases of reionization, and must merge
before the universe achieves the high level of ionization seen
at all redshifts $z < 6$.  Thus, an overlap phase is an inevitable
part of the reionization process.

The spatial distribution of neutral and ionized regions
during reionization can be studied with a wide range of statistics.
Considerable attention has been given already to quantitatively
characterizing the size distribution of ionized bubbles (Furlanetto
et al 2004; Furlanetto \& Oh 2005) and the power spectrum of
the neutral hydrogen distribution, as traced by 21cm emission (see
early work by Madau, Meiksin, \& Rees 1997; Gnedin \& Ostriker 1997; 
Shaver et al 1999; Tozzi et al 2000; and the recent review
by Furlanetto, Oh, \& Briggs 2006).
Here, we suggest the application of a quantitative topological
statistic, the genus number, to the ionized / neutral interface
during reionization.  This will allow us to study properties of
the distribution that are not retained by the other statistics.

Testing topology depends on the dimensionality of the data available. 
Given a two phase medium in three dimensions, it is possible for
either one of the two phases to percolate; this is the ``spongelike''
topology that Gott, Melott, \& Dickinson (1986) predicted for the
cosmic density field. 
In a two-dimensional space, at most one phase can percolate,
while in one dimension, both phases are necessarily broken
into a set of disjoint regions.

Large scale structures have been studied with the three-dimensional
genus curve statistic (Gott, Melott, \& Dickinson 1986 [GMD]; Hamilton,
Gott, \& Weinberg 1986 [HGW]) and its two-dimensional analog (Melott et
al 1989 [M89]).

The conventional Gunn-Peterson test provides a one dimensional
section through the IGM along the line of sight to a bright quasar.
Ionized cavities of size $\ga 1 \pMpc$ in the IGM will
produce sections of spectrum with \lya\ optical depths $\tlya \la 1$,
which can be easily identified.  (Throughout the paper we
denote a physical Mpc by $\pMpc$ and comoving megaparsec by $\cMpc$; 
$1 \pMpc = (1+z) \cMpc$.)
Given enough continuum sources at $z > \zre$, we could
determine the volume fraction of ionized gas as a function of 
redshift.  However, the surface density of suitably bright continuum 
sources is much too low to expect information on multiple sight lines through
any single ionized bubble.

Fortunately, other probes of reionization exist and can more naturally
provide two- and three-dimensional information.  We will focus
primarily on low-luminosity \lya\ emitters as a tool for mapping
neutral gas, but much of the discussion in this paper also applies
to tomographic maps of the interstellar hydrogen made through
redshifted 21 cm line observations (section~\ref{21cm}).

Counts of \lya\ emitting galaxies will be sharply
reduced in a neutral IGM, because \lya\ photons propagating
through neutral gas are resonantly scattered by atomic hydrogen.  
Even the red wing of a \lya\ emission line can be hidden by the 
red damping wing of a substantially neutral IGM (Miralda-Escud\'{e} 
1998; Miralda-Escud\'{e} \& Rees 1998).  This obscures most \lya\ 
photons from view.  While the
resonantly scattered photons will eventually redshift in the Hubble
flow and escape to infinity, their ``photosphere'' is expected to 
subtend $\sim 15''$ (Loeb \& Rybicki 1999), resulting in an essentially
undetectable surface brightness. Most
\lya\ sources detected in a neutral IGM will instead be found using
the modest fraction of \lya\ photons that escape scattering
altogether.  The red wing of the emission line may suffer from a net
optical depth between one and a few (Haiman 2002).  The net reduction 
in line flux is a factor $\ga 5$.  Given the steep \lya\ line luminosity 
function,  this obscuration will
dramatically reduce \lya\ source counts in a neutral IGM.  This method
was first applied to a statistical sample by Rhoads \& Malhotra (2001)
to demonstrate $\zre > 5.7$.  We next applied this test at $z\approx 6.5$ 
(Malhotra \& Rhoads 2004), where we showed that the \lya\ luminosity 
function is consistent with that at $z\approx 5.7$ and inconsistent
with reduction of all \lya\ fluxes by a factor $\sim 3$.  This
implies that the $z\approx 6.5$ IGM is at most $\sim 50\%$ neutral. 
Kashikawa et al (2006) reapplied the test using larger samples.  
They found a deficeit of bright $z\approx 6.5$ galaxies, and suggest
that half the \lya\ may be scattered by the IGM.  
However, other interpretations for this modest reduction 
remain viable, including both true luminosity
function evolution (Dijkstra et al 2006) and field-to-field variations 
in the galaxy number density.  Other \lya\ reionization tests favor
substantial ionized gas at $z\approx 6.5$.  The volume test (Malhotra
\& Rhoads 2006) associates an ionized volume $V_b$ (about 1 pMpc in
radius) with each observed \lya\ source.  The volume neutral fraction $x_n$
is then $\approx \exp(-n_{\lya} V_b)$, where $n_{\lya}$ is the number 
density of \lya\ galaxies.  This yields 
$x_n \la 70\%$ at $z\approx 6.5$.  Tests based on the
spatial correlations of \lya\ galaxies (Furlanetto, Zaldarriaga, 
\& Hernquist 2006; McQuinn et al 2007) work by looking for the 
apparent increase in 
\lya\ galaxy clustering caused by patches of still-neutral IGM.
The first application of these (McQuinn et al) again favors an
ionized IGM at $z=6.5$.

\lya\ emitters can be discovered effectively by at least four
techniques: (1) Narrowband imaging (e.g., Cowie \& Hu 1998,
Rhoads et al 2000, Kudritzki et al 2000, 
Ouchi et al 2001, Malhotra \& Rhoads 2002); 
(2) slitless spectroscopy (e.g., Kurk et al 2004,
Rhoads et al 2005, Pirzkal et al 2007);
(3) slit spectroscopy (e.g., Santos et al 2004, Martin \& Sawicki 
2006); (4) integral field spectroscopy (van Breukelen, Jarvis,
\& Venemans 2005).  Narrowband imaging is the most mature, and 
provides a two-dimensional projection of the \lya\ galaxy 
distribution from a slice of redshift space.  
Narrowband surveys thus provide natural data sets for two-
dimensional genus statistics.  Spectroscopic searches, or complete
spectroscopic followup of narrowband samples, can provide three
dimensional distributions.

The scale of interest
is $\sim 1 \pMpc$,  which is roughly the minimum bubble size that
will render low-luminosity \lya\ sources visible.  At $z\sim 6$, a
$1\%$ bandpass corresponds to a comparable scale along the line of
sight ($\sim 4.5 \Mpc$), while the transverse scale is $\sim 25 \Mpc /
1^\circ$.  Modern large format CCD cameras yield fields of view 
of order $0.5^\circ$, and surveying $\ga 1\sqdeg$ is feasible with
narrowband filters. 
For now, spectroscopic searches yield smaller fields, but a slitless 
grism on a wide field space telescope would be a uniquely powerful 
tool for the type of study we are
considering here.  Such a search might be possible as a byproduct of
several proposed missions.  These include the Joint Dark Energy
Mission concepts DESTINY (Benford \& Lauer 2006) and ADEPT,
and the parallel instrument Mag30Cam for the TPF-C 
mission (Brown et al 2006), all of which plan near-infrared 
grism surveys reaching impressive depths over many square degrees.

\section{Genus Statistics and Reionization} \label{gendesc}
The (3D) genus statistic provides a quantitative measure of the
topology of a two-dimensional surface (typically an isodensity
surface) embedded in a three-dimensional space.  It is heuristically
described as the number of ``donut holes'' in the surface minus the
number of disconnected components.  The two dimensional analog is
the number of closed contours enclosing
high density regions minus the number enclosing low
density regions.   Both are measured by integrating
the curvature of the contour surface and applying the Gauss-Bonnet
theorem, which relates the genus number to this integral (GMD;
HGW; M89).
It is typical to measure the genus number as a function of
the contour threshold, and to parametrize the threshold level 
by the fraction of volume above threshold.  This yields a genus 
curve (HGW; Weinberg, Gott, \& Melott 1987).  When the genus 
measured is for an isodensity surface of a Gaussian random
phase field, HGW show that the genus number per unit 
volume is $G = N (1-\nu^2) \exp(-\nu^2/2)$.  Here
$\nu$ is the contour threshold in standard deviations from 
the mean density, and $N$ depends only on the two-point
correlation function of the field being studied.

The commonest astrophysical application of genus statistics has been
the characterization of density fields, usually traced by galaxies 
(e.g.,  Gott et al 1989) and/or galaxy clusters (e.g., Rhoads, Gott \&
Postman 1994).  Additional applications of the 2D statistic
have been made or proposed for microwave background maps and for
gravitational lensing shear measurements (e.g., Jain \& Matsubara 2001).

There is also a one-dimensional analog, the threshold crossing statistic,
which has been applied to quasar spectra at $z\approx 6$ (Fan et al 2002).

\section{Requirements for a Lyman $\alpha$ Topological Test of Reionization}
\label{lya_req}
To successfully identify the topological signature of \ss\
overlap in a search for \lya\ emitters, a few conditions must be
fulfilled.  First, the typical \ss\ radius must be large enough that a
substantial fraction of flux escapes from \lya\ emitters in the
bubble.  The optical depth of \lya\ damping wing absorption 
from neutral gas outside a \ss\ is 
\begin{equation}
\tau \approx {1.2 \pMpc \over r_p + \Delta v / H(z)}
 = {1.2 (1+z) \cMpc \over  r_c + (1+z) \Delta v / H(z)}
\end{equation}
for \lya\ photons emitted at a velocity offset $\Delta v$ from the
systemic velocity of a galaxy located a physical distance $r_p$ (or
comoving distance $r_c \equiv (1+z) r_p$) from the edge
of the \ss.

A bubble radius of $1.2 \pMpc = 1.2 (1+z) \cMpc$ results in a line
center optical depth $\tau = 1$ (e.g., Rhoads \& Malhotra 2001). 
Bubbles of this size or larger will be apparent in the observed
distribution of \lya\ galaxies, while substantially smaller bubbles
will not.  
The characteristic bubble size, based on a Press-Schechter style
model, goes from $\sim 2\, \cMpc$ to $\sim 20\, \cMpc$
as the volume neutral fraction drops from $0.7$ down to 
$0.26$ (Furlanetto et al 2004).  Thus, the bubbles are large
enough to allow substantial transmission of \lya\ around the epoch
of 50\% neutral fraction.

Second, each bubble must contain at least a few detectable
\lya\ sources.
Like the first constraint,
this one is basically a requirement on the typical \ss\ size.
\lya\ luminosity functions are typically modeled as Schechter functions, 
$\Phi(L) dL = \Phi^* (L/L^*)^\alpha \exp(-L/L^*) dL / L^*.$
If we consider a survey with luminosity threshold $L_{min}$, the
resulting requirement is 
\begin{equation}
V_b \, \int_{L_{min}}^\infty \phi(L) dL = {4 \pi\over 3} \, r_b^3 \Phi^* 
\, \Gamma(\alpha+1, L_{min}/L^*) \ga 4 ~~.
\label{nlya}
\end{equation}
Here $\Gamma$ is the upper incomplete Gamma function, $r_b$ and $V_b$ are
the characteristic bubble radius and volume, and 4 is a
working minimum number of sources per bubble for genus curve
measurements. 
A $\sim 1.2 \pMpc$ bubble (cf. eq.~1)  has volume 
$2.5\times 10^3 \left((1+z)/7\right)^3 \cMpc^3$.
To meet condition~\ref{nlya}, then, we would
like $\ga 2\times 10^{-3} {\hbox{ galaxies }} \cMpc^{-3}$ above 
a survey's detection threshold.
Observed \lya\ luminosity functions at $z \approx 6.5$ are 
fit by $\alpha = -1.5$,  $L^* \approx 10^{42.6} \erg$ 
(Malhotra \& Rhoads 2004; Kashikawa et al 2006) and $\Phi^*$ 
from $\approx 10^{-3.3} \cMpc^{-3}$ (MR04) to $\approx 10^{-2.88} 
\cMpc^{-3}$ (Kashikawa et al 06).  The desired number density is
then achieved for $L_{min} \approx 10^{41.5}$--$10^{42} \ergsec$.
A rise in the number density of faint \lya\ galaxies at $z\sim 10$ has 
recently been suggested (Stark et al 2007), which would help with
condition~\ref{nlya} if confirmed.

Third, as shown by Gott and collaborators (GMD, HGW, M89), galaxy
distributions have their own characteristic topological signatures.
Before overlap,  the topological signature of
the neutral and ionized regions may be strong enough to completely
dominate the topology of \lya\ emitters.  During and after overlap,
however, this is unlikely to remain true.  By using
a control sample of continuum-selected galaxies alongside the 
\lya\ sample, we can separate the topological signatures
of the intergalactic medium from those of the galaxy distribution
itself.  This could be accomplished through a deep Lyman break
galaxy survey.  The smoothed \lya\ galaxy density field would
then be divided by the smoothed density field of continuum-selected
galaxies.  Such a method would work best if the continuum sample
has a higher volume density than the \lya\ selected sample;
otherwise, the uncertainties in the experiment would increase
quite substantially.

Of course, if these conditions are not all fulfilled, a topological
study of the large scale \lya\ source distribution will still reveal
interesting information about galaxy formation and reionization.  In
particular, if the topological signature of isolated bubbles is not
observed at any redshift, it will indicate that the sources of
ionizing radiation are of low luminosity and the bubbles they produce
are consequently small.

\begin{figure}
\epsscale{1.0}
\plotone{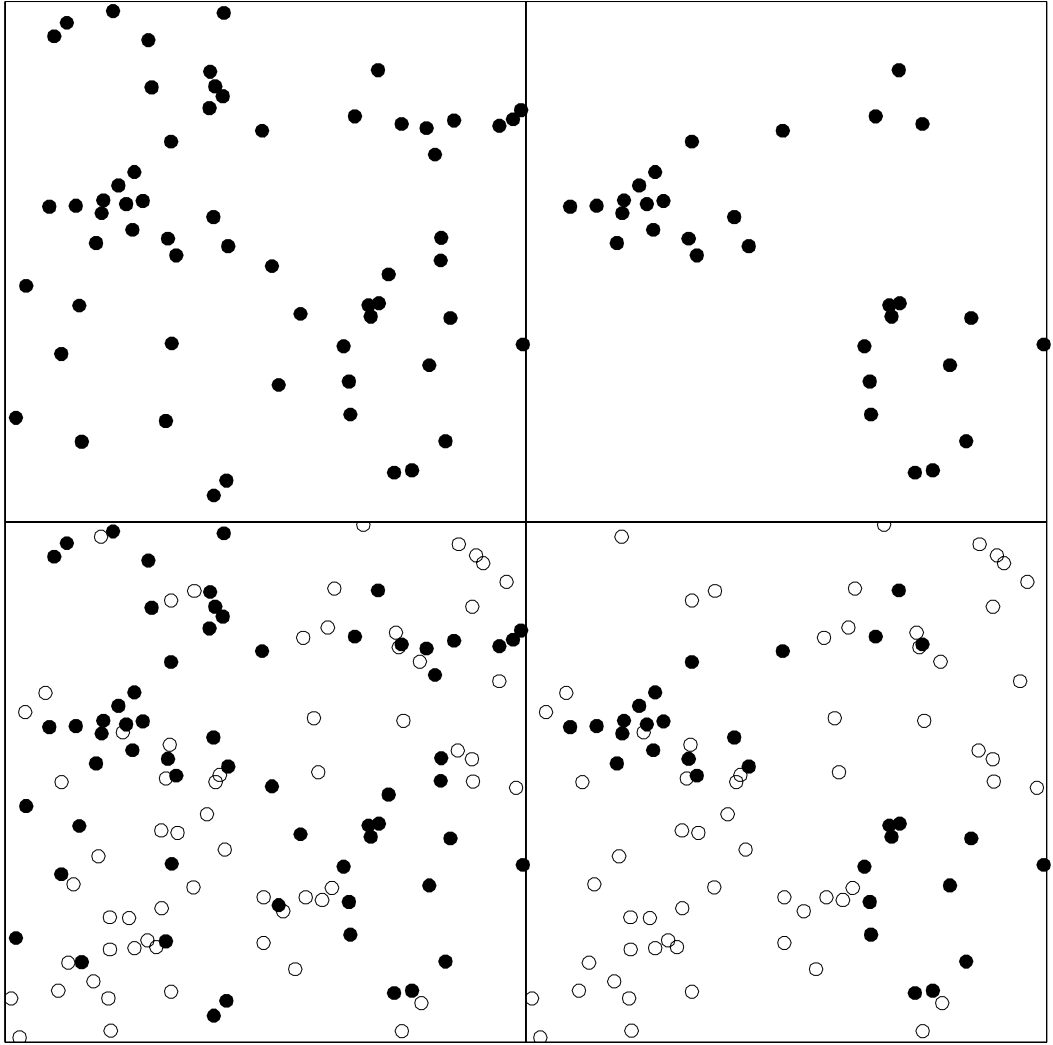}
\caption{The conceptual basis for mapping ionized bubbles
using \lya\ galaxies.  (a) Top left: A hypothetical projected
distribution of \lya\ galaxies, shown by filled circles. 
(b) Top right: The same distribution
as it might be observed during the central phase of reionization
using a line-selected sample. (c) Bottom left: As in (a) but with 
a hypothetical Lyman-break selected sample shown by open circles.
(d) Bottom right: As in (b), but again showing the break-selected
control sample.
Panel (b) shows increased clumping and reduced
number density compared to (a).  However, unambiguous 
interpretation would be much easier with a control sample
selected by some method independent of the IGM properties,
as shown in panel (d).
\label{bub_cartoon}}
\end{figure}

\section{The Genus Curve of the Ionized-Neutral Interface}
The genus number for the neutral-ionized interface can be predicted as
a function of the ionized volume fraction based on simple physical
arguments that describe the way reionization progresses.  Formally, we
define this ``interface'' as a fixed contour
level in the local neutral fraction, $x_{thr} \sim 0.5$.  So long as
the boundaries of HII regions are thin, as expected when 
reionization is driven primarily by stellar photons,
the exact choice of $x_{thr}$ matters little.

After reionization begins but well before overlap, the IGM
consists of a collection of isolated, ionized bubbles embedded
in a matrix of neutral gas.  Each bubble may correspond to
a single ionizing source, or to the ensemble of all ionizing sources
in a particular peak of the density field (FZH; FO),
or something in between.  In this regime, the genus number
of the neutral-ionized interface per unit volume 
is simply $-1$ times the number density of (surviving) ionized
bubbles.  Of course, any
practical measurement will depend on the ionized regions achieving
some minimum detectable size, and the relevant number density should
really be measured above this minimum size.
The size will be closely related to the total ionizing photon production
of the bubble, given by $L_i \times t$, the product of
ionizing luminosity and source age or lifetime. 
However, it will also depend on the local gas density.

Once bubbles grow beyond the immediate neighborhood of the ionizing
sources, the likely pattern of reionization depends on details of the
model.  Miralda-Escud'{e} et al (2000) used a semi-analytic approach
based on a statistical description of the IGM density distribution.  They
argued that ionizing photon production is substantially balanced by
recombinations.  Because the recombination rate per ion scales with
the density of ionized gas, they conclude that the low-density regions
will ionize first, and that a large fraction of ionizing photon production
will be balanced by recombinations at the neutral-ionized interface
where the ionized gas is densest.  As time goes by and the ionizing
photon production rate rises, reionization gradually proceeds to the
highest density regions.

More recently, studies based on numerical simulations (e.g., Iliev et
al 2006) have concluded that reionization proceeds from the highest
density regions, where the ionizing photon production is strongly
concentrated, to the lowest density regions.  In this picture, the
clustering of ionizing sources is more important than the density
variations in intergalactic gas.

Fortunately, whether reionization proceeds ``inside out'' or ``outside
in,'' the neutral-ionized interface will approximately follow the
cosmic density field.  This reduces the problem to the well-studied
case of the genus number of the galaxy distribution in the linear (or
nearly linear) regime. Because the genus curve of the density field is
itself symmetric for gaussian random phase initial conditions, we
expect the genus number of the neutral-ionized interface to behave
similarly with volume ionized fraction in either scenario.  In the
central phases of reionization, the 3D genus number of the
interface will become positive, reflecting the
transition to a multiply-connected topology for both the ionized and
neutral phases of the IGM.

The amplitude of the neutral-ionized interface genus number 
can be estimated from two characteristic lengths: The 
typical bubble size $\langle r_b \rangle$, and the effective smoothing scale 
$\lambda_s$ of the observations.  For \lya\ galaxy tests, 
radiative tranfer implies a line-of-sight smoothing with 
$\lambda_s \approx 1.2 \pMpc$.  The requirements discussed
in section~\ref{lya_req} suggest a comparable smoothing length
should be applied in the transverse directions.  Then the expected
genus number density becomes $\sim (2\pi)^{-2} \lambda_s^{-3} \sim
10^{-5} \cMpc^{-3}$ (see, e.g., eq.~8 of Rhoads, Gott,
\& Postman 1994).  Since modern narrowband surveys with half-degree
cameras cover volumes $\sim 10^5 \cMpc^{3}$ per field, 
a \lya\ survey of a few contiguous fields should contain enough
volume for an interesting measurement of the neutral-ionized
interface genus number.  If $\langle r_b \rangle \gg \lambda_s$, we
would need to substitute $\langle r_b \rangle$ for $\lambda_s$ in our
estimated genus number density.

\begin{figure}
\epsscale{1.0}
\plotone{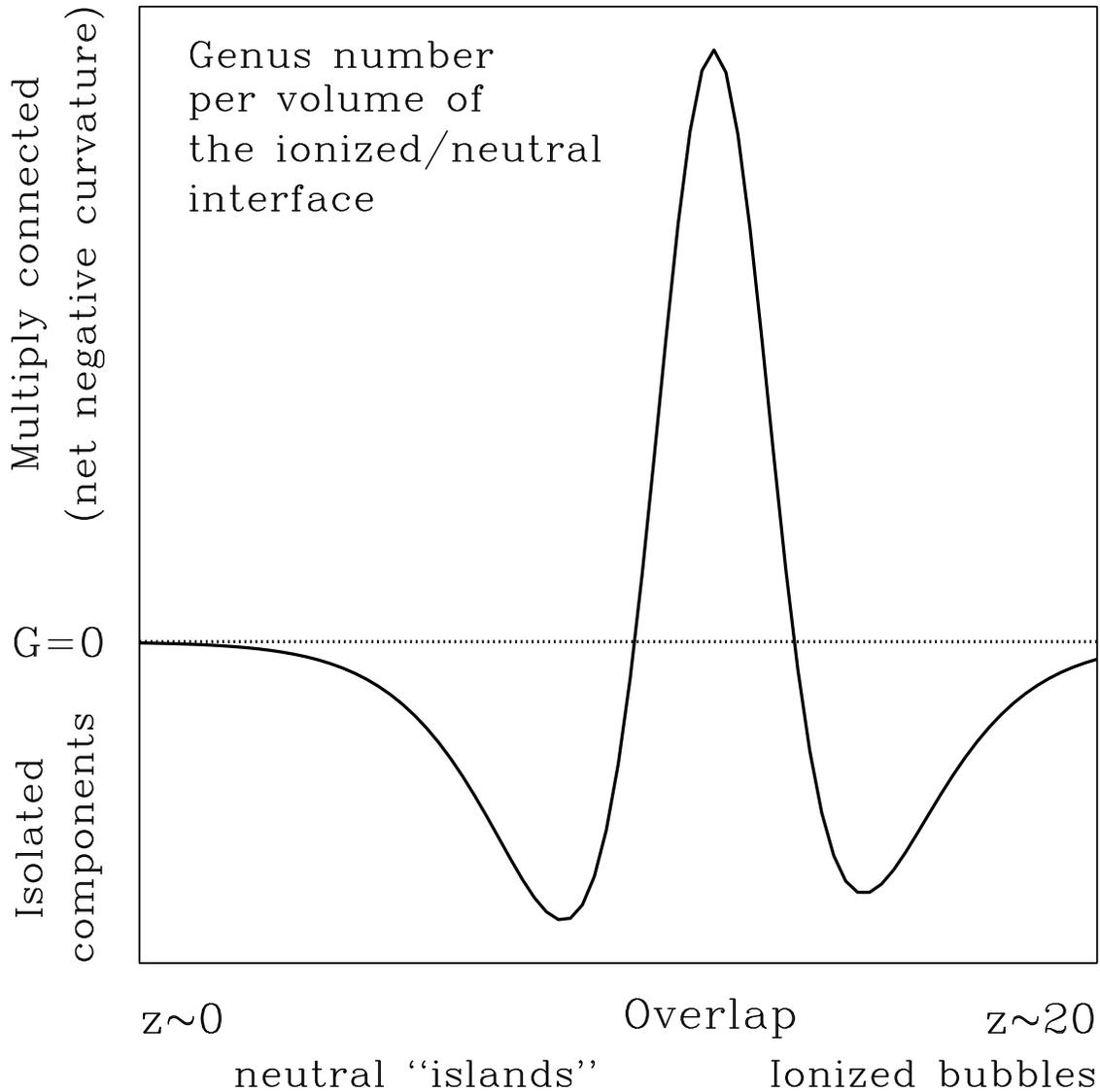}
\caption{Schematic illustration of the genus number per unit
volume of the neutral/ionized interface during reionization.
This surface consists of disconnected components during early
reionization, when isolated bubbles of ionized gas grow 
in a neutral universe.  This corresponds to a net negative
genus number.  The sign of the genus number changes during
the overlap phase, when the ionized region begins to percolate.
At this stage, both neutral and ionized phases of the gas
are multiply connected, and the interface surface has net
negative curvature, corresponding to a positive genus number.
Finally, at late time, only isolated ``islands'' of neutral 
gas remain. The interface surface is again a collection of 
disconnected components, and the genus number is again negative.
The expected amplitude of the curve is $\sim 10^{-5} \cMpc^{-3}$ 
for \lya\ galaxy based tests (see text).  It could in principle 
be much higher for 21cm based tests that resolve smaller ionized 
bubbles.
\label{gen_cartoon}}
\end{figure}

A sketch of the expected genus evolution is shown
in figure~\ref{gen_cartoon}. A detailed prediction for the magnitude of
the genus number and its evolution with redshift depends 
on the many unknown details of the reionization process (feedback
effects, possible dependence of the escape fraction on galaxy
mass, etc), so we do not attempt a detailed analytical treatment
here.  The best way to make model the genus number in this
phase will be genus curve ``measurements'' in reionization simulations,
preferably coupled with a realistic algorithm for assigning 
\lya\ galaxies to dark matter halos.

\section{The 21 cm Line and the Topology of Reionization} \label{21cm}
The 21 cm emission line from the hyperfine transition of neutral
hydrogen offers a complementary probe of ionized region topology.
Sufficiently sensitive 21 cm line maps would
allow much smaller ionized bubbles to be measured, for two
reasons. First, there is no discreteness in the HI emission,
unlike the \lya\ emitters, and so there is no corresponding
requirement that an identifiable bubble be large enough to contain
multiple sources.  Second, the typical velocity resolution of radio
telescopes is a few $\kms$.  Bubbles whose expansion
velocity $r H$ is larger than this should be visible to sufficiently
sensitive 21 cm mapping.  This is much smaller than the minimum bubble
expansion velocity scale for the \lya\ test, which may be estimated as
$\sim 1.2\Mpc H \approx 1000 [(1+z)/8]^{3/2} \kms$ (see
section~\ref{lya_req}).

A more fundamental limit to the resolution of 21 cm mapping along the
line of sight is set by the peculiar motions of the IGM.  The
velocities within ionized gas bubbles are not relevant, of course,
since the emission comes only from the neutral regions.  The velocity
in the neutral regions is unlikely to be affected by 
nongravitational physics at any level much greater than $\sim 50
\kms$, which is the velocity corresponding to the ionization potential
of hydrogen.  Gravitationally driven velocities are similarly unlikely
to much exceed tens of $\kms$ outside virialized halos, which
consititute a very small fraction of baryons at the epoch of
reionization.  Thus, 21 cm mapping could identify
bubbles down to at least $\sim 50 \kpc$ proper size ($\sim 0.5 \Mpc$
comoving), and probably smaller, provided sensitivity and foreground
contamination issues can be overcome.

The primary concern in 21 cm reionization tests is
contamination from foreground radio sources and from free-free
emission in the ionized bubbles at the epoch of reionization.  The 
best strategy for removing both foregrounds may be mapping in
frequency space, because the contaminating signals vary as
featureless power laws on velocity scales up to several 
thousand $\kms$.  Still, the foregrounds are dramatically brighter 
than the reionization signals, and exquisite foreground subtraction 
is needed.
Current 21cm experiments (e.g., LOFAR [Falcke et al 2006], the Mileura
Widefield Array [MWA; Morales et al 2006], and the Primeval Structure
Telescope [PAST; Peterson, Pen, \& Wu 2005]) are all aiming first for
statistical detections of the 21cm reionization signature, with bubble
mapping a possible future step only if radio interference and
astronomical foregrounds prove sufficiently tractable.

\section{Summary and Future Outlook}
The overlap phase of reionization can be defined topologically
and sought using a topological statistic, the genus number.
By doing so, we can identify the central period of the reionization
process.  The topological signature of overlap will work whether
reionization proceeds from high- to low-density regions or
vice versa.  Topological tests could be applied using any
method that allows mapping ionized and neutral regions during
the epoch of reionization.  Both 21cm emission from neutral gas
and obscuration of \lya\ galaxies by neutral gas could provide
such maps. Each tool has its own strengths: The 21cm maps can in 
principle map smaller bubbles, but difficulties associated
with foreground removal will not be fully resolved before
the first test data from new experiments becomes available.

We have given particular attention to applications of this 
test based on mapping the distribution of \lya\ emitting galaxies.
For reasonable expectations, the ionized bubbles during overlap 
will be large enough to permit transmission of \lya\ flux through
an otherwise neutral medium.  Moreover, the number density of \lya\ 
emitting galaxies provides a sample of several objects per bubble. 
Complementary surveys for Lyman break selected galaxies can provide
control samples to separate the topological signature of structure
formation from that of reionization.

Applications of this test are practical at redshifts up to $6.9$ with
CCD cameras, and at least to $z\sim 10$ with the new generation of
large format near-infrared detectors now becoming available.  Indeed,
several surveys for \lya\ galaxies at $z\ga 8$ are already underway.
Thus, \lya\ galaxies could
well provide our first direct evidence for individual ionized bubbles
in the epoch of reionization.  Topological tests for the overlap phase
would be practical once a contiguous volume corresponding to several
bubbles had been mapped.

\acknowledgements 
I thank Steven Furlanetto, Benedetta Ciardi, Michael Strauss, and 
Sangeeta Malhotra for helpful discussions on various aspects
of this idea over the last few years.

\end{document}